# Agentic Artificial Intelligence for Reproducible Human-in-the-Loop Environmental Health Research

Edmund Seto[1*]

1. Department of Environmental & Occupational Health Sciences
   School of Public Health
   University of Washington
   Seattle, WA

* Correspondence: Edmund Seto, PhD
Professor
Email: eseto@uw.edu

## Abstract

Agentic artificial intelligence (AI) systems that are capable of planning and executing multi-step analytical tasks are increasingly available to environmental health researchers, but their reliability in real-world practice has not been fully explored. This paper describes a human-in-the-loop agentic framework for environmental health research, involving the review, verification, and correction of AI-generated data analysis code and results at each step – a process that mirrors the mentorship structure of traditional research teams. This approach offers a path toward rigorous, reproducible use of agentic AI in routine data-rich environmental health research. We illustrate this framework through a case study analyzing nitrogenous organic contaminants at U.S. Superfund sites, a chemical family linked to the emerging tire-derived contaminants 6PPD and 6PPD-quinone. Using an agentic large language model to generate R code for data filtering, spatial mapping, and cluster analysis, we document instances where initial agentic AI outputs benefited from a human-in-the-loop process to produce more rigorous and reproducible results. We conclude that effective use of agentic AI requires both domain expertise to frame questions and evaluate outputs, and coding literacy to guide the AI's approach, while outlining future opportunities for agentic AI workflow to advance the environmental health sciences.

## Keywords

Agentic artificial intelligence, Human-in-the-loop, Large language models, Reproducibility, Environmental health, Superfund, Nitrogenous organic contaminants, 6PPD-quinone

## 1. Introduction

### 1.1 The History of Data Science and Artificial Intelligence in Scientific Research

Scientific inquiry has historically progressed through recognizable paradigms: periods driven by empirical observation from experiments, theoretical frameworks built on formal deduction, and, over the past several decades, computational modeling. Many historians of science now describe a fourth, data-driven paradigm, in which discovery is increasingly mediated by the integration of large-scale data, high-performance computation, and, most recently, artificial intelligence (AI) (1). The modern intellectual roots of AI trace back to Alan Turing's formalization of computability and the 1956 proposal by McCarthy et al., which coined the term (2–4). For much of the following half-century, AI and data analysis progressed incrementally. More recently, however, the convergence of expanding open scientific data, inexpensive computing power, and large statistical models has produced a qualitative shift in the pace and scale of AI-enabled discovery across the sciences. Data science methods, once the province of computer science and statistics, are now embedded throughout genomics, geospatial science, clinical research, and environmental health.

### 1.2 Artificial Intelligence and Machine Learning in Environmental Health Sciences

Environmental health science is a natural beneficiary of this shift. The field relies on integrating heterogeneous, high-dimensional data – including chemical structures, toxicity profiles, remote sensing, geospatial exposure surfaces, electronic health records, sensor streams, and biomedical literature – to characterize how environmental exposures impact human health. Machine learning (ML) and AI methods are increasingly used to address this integration challenge. Recent peer-reviewed work illustrates the breadth of these applications: ensemble and deep-learning models now support chemical toxicity and quantitative structure-activity relationship (QSAR/QSPR) predictions, Explainable AI (XAI) methods are being applied to interpret "*black box*" toxicological models for regulatory use, and ML is used to generate high-resolution spatiotemporal exposure surfaces (e.g., for fine particulate matter) that were previously unattainable from sparse monitoring networks alone (5). Geospatial AI (GeoAI) methods, which combine ML with geographic information systems (GIS) and increasingly rich sensor and satellite data, enable scalable exposure assessment in large population cohorts and administrative health databases (6).

Within the universe of AI technologies, large language models (LLMs) are entering environmental health practice for literature synthesis, structured data extraction, hypothesis generation, and preliminary environmental decision support (7–9). At the same time, existing literature is candid about the technology's limitations and risks, including hallucination, opacity, bias, and the substantial computational and environmental footprint required to train and run large models. These challenges underscore the need for expert-supervised deployment rather than unchecked automation (9).

The National Institute of Environmental Health Sciences (NIEHS) has actively promoted these methods. At the September 2023 meeting of the National Advisory Environmental Health Sciences Council, NIEHS leadership and invited experts described AI as central to the future of the discipline, highlighting its utility for integrating toxicology data across chemical mixtures, making sense of wearable sensor data, and mapping the exposome across the lifecourse (10). More recently, NIEHS highlighted investments by the NIH Office of Data Science Strategy in cloud computing infrastructure, data-sharing standards, and workforce training designed to help environmental health investigators use AI and ML responsibly (11). NIEHS has also emphasized the ethical dimensions of this transition, including the necessity of guidelines for AI-generated synthetic data (12). Taken together, institutional momentum and expanding scientific literature demonstrate that AI and open data science are no longer peripheral tools for environmental health research but are becoming fundamental components of its core methodological infrastructure.

### 1.3 The Evolution of Artificial Intelligence: From Symbolic Systems to Agentic, Human-Supervised Frameworks

To situate agentic AI and human-in-the-loop methods within this broader scientific context, it is useful to trace the evolution of AI chronologically, in five overlapping phases.

Phase 1: Symbolic and Rule-Based AI (1956–1980s). Systems in this era encoded expert knowledge as explicit logical rules, as exemplified by early expert systems. Performance depended entirely on the completeness of hand-crafted rule sets, and these systems lacked the ability to learn from data (4).

Phase 2: Statistical ML (Late 1980s–2000s). Rather than relying on hand-coded rules, algorithms such as decision trees, support vector machines, and random forests were trained to detect patterns directly from labeled data. This era enabled applications like the QSAR toxicity models still widely used in environmental health today (5).

Phase 3: Deep Learning Revolution (Post-2012). Multi-layer neural networks trained on large datasets using graphics processing units (GPUs) began to outperform prior methods on tasks such as image classification. This breakthrough laid the groundwork for applications like automated analysis of remote-sensing and medical-imaging data in exposure and health research.

Phase 4: Transformer Architecture and LLMs (Post-2017). The introduction of the transformer neural network architecture enabled efficient training on vast text corpora, giving rise to modern LLMs (13). Successive generations demonstrated general-purpose capabilities in language understanding, reasoning, and code generation – skills directly applicable to scientific data analysis (14).

Phase 5: Agentic AI and Human-in-the-Loop Integration (Present). The current phase centers on how LLMs interact with external tools and each other. Two related developments define this era: First, the Model Context Protocol (MCP), an open standard released by Anthropic in 2024 for connecting AI models to external tools, databases, and files through a common interface, addressed a long-standing bottleneck in which every data source or software tool required a custom integration (15). Building on this connectivity, agentic AI systems do not simply answer a single question, but can plan, sequence, and execute multi-step tasks. They can call external tools, write and execute code, and refine their approach based on intermediate outputs to accomplish the greater objective (14,16).

Because autonomous, multi-step AI actions introduce new risks, ranging from factual errors and hallucinations to opaque decision-making, a parallel methodological literature has emphasized "*human-in-the-loop*" (HITL) AI. In HITL frameworks, human judgment is deliberately embedded at defined checkpoints in an automated pipeline (17). Systematic reviews demonstrate that combining human oversight with automated execution consistently outperforms either humans or AI systems working in isolation on complex, high-stakes tasks. This synthesis of an LLM capable of agentic, tool-using behavior disciplined by structured human review forms the methodological foundation for the framework presented here.

### 1.4 Traditional Team Science Aligns with Human-in-the-Loop Agentic AI

The HITL structure described above is not a novel invention; rather, it closely mirrors the long-standing organizational structure of academic research teams. In the traditional model, an undergraduate, graduate student, or postdoctoral researcher conducts analyses, writes code, organizes the results, and subsequently reports to a senior investigator, who reviews, critiques, and redirects the work before it advances (18). Postdoctoral researchers, in particular, often produce independent work while simultaneously supervising junior trainees, creating a layered chain of iterative review that extends from undergraduate students to a principal investigator. Scientific progress in this model is rarely linear; it advances through repeated cycles, feedback, correction, and refinement, with experienced team members identifying errors, challenging assumptions, and ensuring conclusions are rigorously supported.

This structure offers a productive analogy for agentic AI with HITL checkpoints (Figure 1). In this framework, an AI agent operates in a role analogous to a junior trainee: it can rapidly execute well-defined analytical tasks, such as querying databases, cleaning and transforming datasets, or drafting statistical code, but its outputs are not treated as final. Instead, intermediate outputs are routed to the human investigator, who acts as the senior mentor to review, correct, and approve the work before proceeding to subsequent stages. As in a traditional research group, oversight is not a single post-hoc evaluation at project completion, but rather a series of checkpoints distributed throughout the analytical workflow.

This analogy is imperfect though: unlike a human trainee, an AI agent lacks independent judgement and accountability, and without computational equivalents of long-term memory or continuous learning, it does not accumulate experience over time. Consequently, the human investigator retains complete responsibility for the scientific and ethical integrity of the research. Nonetheless, the underlying logic, that complex, consequential work benefits from being decomposed into stages separated by review and refinement, is precisely the principle that has long governed how research groups safeguard scientific quality, and it is this principle that HITL agentic AI operationalizes computationally.

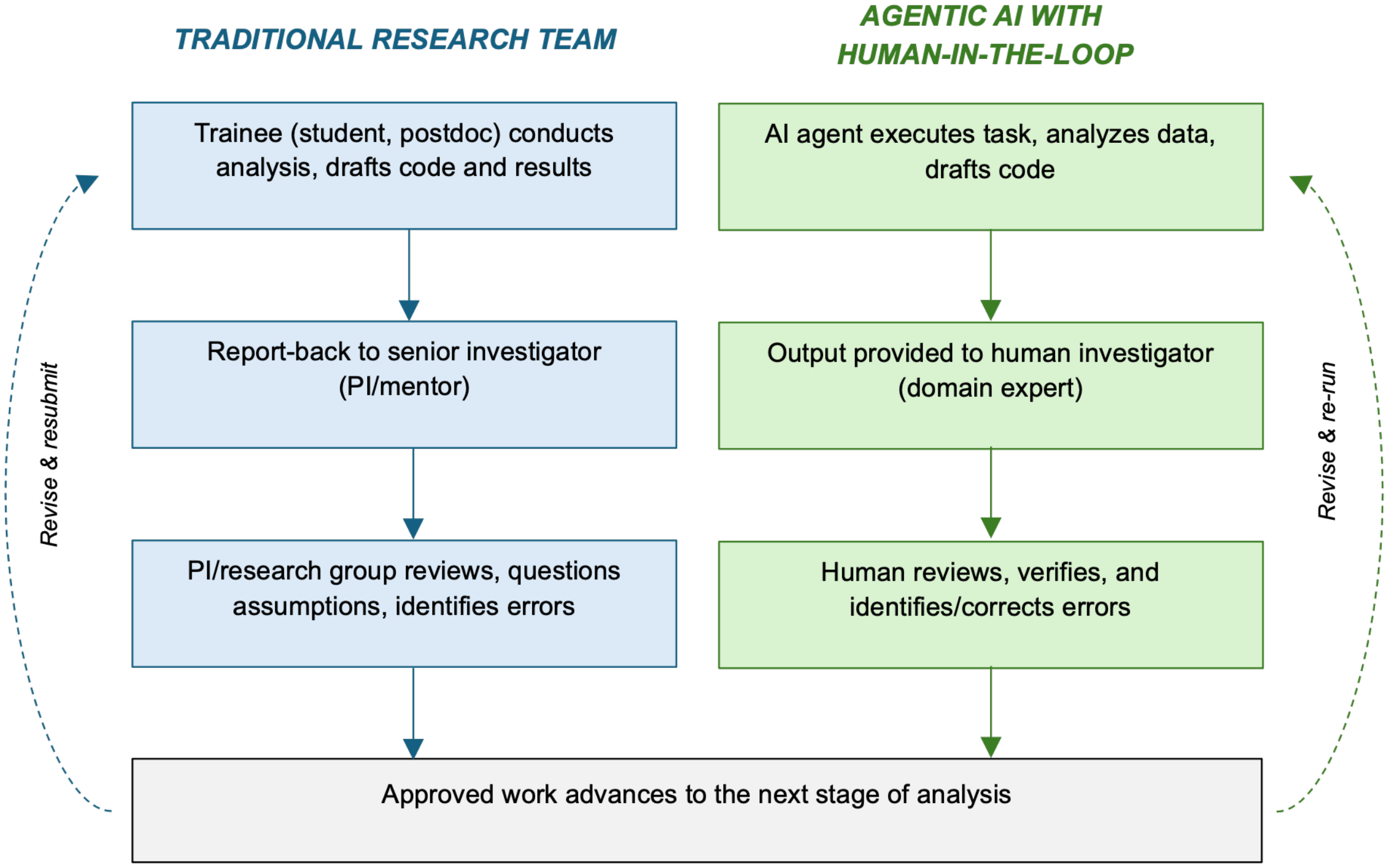

**Figure 1.** Human-in-the-loop agentic AI as a computational analogue of traditional research team mentorship. In both models, an initial work product, drafted by a trainee or executed by an AI agent, is routed to an experienced reviewer(s) for verification and correction before advancing, with the review cycle repeated at each stage of the analytical workflow.

### 1.5 Scientific Rigor and Reproducibility in the Age of Agentic AI

Scientific rigor refers to the strict application of the scientific method to ensure robust and unbiased experimental design, execution, analysis, and interpretation. Reproducibility, narrowly defined, is the ability of an independent analyst to obtain identical results using the original data and code (19). Reproducibility is distinct from, but foundational to, replicability, which is the ability to reach consistent conclusions using new data collected under comparable conditions. If a result cannot even be reproduced from the original data, its replicability across new datasets cannot be meaningfully evaluated.

The importance of rigor and reproducibility has been underscored by a substantial body of evidence documenting widespread failures. Ioannidis argued on statistical and methodological grounds that a large share of published research findings across biomedicine may be false (20). A large-scale replication effort by the Open Science Collaboration attempted to reproduce 100 published psychology studies and found that only 36% yielded statistically significant results upon replication, compared to 97% of the original studies (21). Similarly, a 2016 Nature survey of over 1,500 scientists found that more than 70% had failed to reproduce another researcher's results at least once, and over half had failed to reproduce their own prior findings (22). These findings, termed the "*reproducibility crisis*," have prompted calls across scientific disciplines, including environmental health, for practices that make the full analytical pipeline transparent, verifiable, and shareable (19).

Agentic AI with HITL oversight offers an opportunity to advance these goals for two key reasons. First, modern agentic AI systems built on coding-capable LLMs perform data analysis by generating inspectable code (e.g., in Python or R) rather than opaque numerical outputs. As a result, every stage – from data cleaning and statistical modeling to data visualization – exists as a discrete, human-readable artifact that can be evaluated, rerun, version-controlled, and audited in alignment with standard computational research guidelines (19). Second, HITL checkpoints enforce the exact expert verification that reproducibility scholars consider essential, but which is sometimes skipped under time and resource constraints. Domain experts can confirm that each analytical step is scientifically sound before the workflow advances, avoiding the post-hoc discovery of errors after manuscript preparation or publication (19,20). In this sense, the combination of AI-generated, executable analytical code with structured human review does not merely automate data processing; it embeds transparency and verification directly into the scientific process, in a form that can be archived and shared alongside the research publication.

## 2. A Case Study: Nitrogenous Organic Contaminants at U.S. Superfund Sites

### 2.1 The Superfund Program, the National Priorities List, and the Substance Priority List

To demonstrate the practical application of a HITL agentic AI framework, we present a data analysis case study driven by stakeholder concerns regarding a specific class of emerging and less-understood chemicals found across U.S. Superfund sites.

Congress established the Superfund program through the Comprehensive Environmental Response, Compensation, and Liability Act (CERCLA) of 1980, granting the U.S. Environmental Protection Agency (EPA) authority to identify, investigate, and remediate sites contaminated with hazardous substances (23). Candidate sites are evaluated using the Hazard Ranking System (HRS), a numerical scoring tool that assesses a site's potential threat to human health and the environment based on data from preliminary assessments and site inspections. This evaluation incorporates exposure pathways, waste characteristics, and affected populations or ecosystems (24). Sites scoring above a statutory threshold are proposed for the National Priorities List (NPL) and, following public comment, formally added. The NPL serves as EPA's list of the nation's most contaminated sites and provides the mechanism by which sites become eligible for federally funded long-term cleanup actions (25).

A related but distinct list, the Substance Priority List (SPL), is prepared jointly by the Agency for Toxic Substances and Disease Registry (ATSDR) and EPA under CERCLA Section 104(i), as amended by the Superfund Amendments and Reauthorization Act. Rather than ranking sites, the SPL ranks approximately 275 hazardous substances detected at NPL sites using a composite score based on three criteria: frequency of occurrence at NPL sites, toxicity, and potential for human exposure (26). Because the algorithm weights frequency and exposure potential alongside toxicity, ATSDR explicitly notes that the SPL is not merely a list of the "*most toxic*" substances; a substance with moderate toxicity can rank highly if it occurs frequently at NPL sites and human exposure potential is high. The list is updated roughly every two years. Together, the NPL and SPL provide two complementary priority-setting tools: the NPL identifies which contaminated locations warrant federal attention, and the SPL identifies which contaminants, wherever they occur among those places, warrant the most attention from a public health perspective.

### 2.2 Nitrogenous Organic Contaminants of Concern: A Multi-Class Chemical Family

The nitrogenous organic chemicals prioritized in this case study are not a single narrow class. Instead, they span several structurally distinct, yet toxicologically related subclasses unified by a nitrogen-bearing functional group attached to, or derived from, an aromatic ring system.

The most direct subclass is the aromatic amines (arylamines): primary and secondary anilines in which an amine group is bonded to a benzene ring. Examples include aniline, benzidine, 3,3′-dichlorobenzidine, 4-aminobiphenyl, 4,4′-methylenebis(2-chloroaniline) (MBOCA), and the secondary aromatic amine diphenylamine. This subclass has a long industrial history as intermediates in the manufacture of synthetic dyes, pigments, rubber chemicals, and pharmaceuticals, and includes several IARC Group 1 and 2A human carcinogens implicated in occupational bladder cancer among dye, rubber, and textile workers (27,28).

A second, closely related subclass comprises hydrazines and azo compounds, represented here by 1,2-diphenylhydrazine (hydrazobenzene) and its oxidation product azobenzene. Rather than an unrelated chemical family, these represent a masked or precursor form of the aromatic-amine hazard. Azo dyes, one of the largest classes of synthetic dyes manufactured worldwide, are built around the same –N=N– azo linkage present in azobenzene. Reductive cleavage of that bond, whether during wastewater treatment, in the environment, or in the human gut, regenerates free aromatic amines such as benzidine and other regulated carcinogens (29).

A third subclass, nitroaromatic compounds, replaces the amine group with a nitro ($–NO_2$) substituent on the aromatic ring, as in nitrobenzene, 4-nitroaniline, and 4-nitrophenol. Nitroaromatics share industrial origins with the aromatic amines, as aniline itself is manufactured by the reduction of nitrobenzene. Furthermore, they act through related toxicological mechanisms, including redox cycling and reactive oxygen species generation implicated in mutagenicity and carcinogenicity (30).

A fourth subclass, *N*-nitrosamines, includes *N*-nitrosodimethylamine (NDMA), *N*-nitrosodiphenylamine, and *N*-nitrosodi-*n*-propylamine. Nitrosamines form when a secondary or tertiary amine reacts with a nitrosating agent, and several, including NDMA, are among the most potent DNA-reactive carcinogens identified in rodent bioassays. *N*-Nitrosodiphenylamine is the direct nitrosation product of diphenylamine and was historically used as a rubber vulcanization retarder, tying this subclass directly to the rubber-manufacturing sector (31).

Finally, carbazole represents a nitrogen-containing heterocyclic aromatic compound structurally related to, and typically co-occurring with, polycyclic aromatic hydrocarbons. It is a minor but environmentally persistent constituent of coal tar and creosote, as well as a recurrent contaminant in groundwater at coal-tar and wood-treatment NPL sites (32). This connection is directly relevant to the Pacific Northwest, where former creosote wood-treating facilities are among the region's most prominent NPL sites. These include the Wyckoff Company/Eagle Harbor site on Bainbridge Island, Washington, and the McCormick & Baxter Creosoting Company site in Portland, Oregon, both listed on the NPL after decades of creosote-based wood treatment contaminated adjacent soil, groundwater, and sediments (33,34).

Despite this structural diversity, these subclasses converge on a common toxicological theme and a shared industrial lineage. Toxicologically, each subclass is metabolically or environmentally activated to a DNA- or protein-reactive electrophilic intermediate: aromatic amines to hydroxylamine and nitrenium-ion species, azo compounds to those same aromatic amines upon reductive cleavage, nitroaromatics to nitro-radical anions and reactive oxygen species, and nitrosamines to alkylating diazonium intermediates (28–30). Industrially, several subclasses trace to the same dye, chemical, and rubber-manufacturing sectors, explaining why members of more than one subclass frequently co-occur as contaminants of concern at NPL sites, and why occupational exposure to this broader chemical family has long been associated with elevated cancer risk (28,31). Beyond human health, members of this class pose ecological risks: several exhibit acute or chronic toxicity to aquatic organisms, display highly mobility and persistence in groundwater, and, in the case of carbazole, persist in aquifers as recalcitrant residuals long after more degradable coal-tar constituents have attenuated (27,32).

### 2.3 From Legacy Contaminants to an Emerging Contaminant Class: 6PPD, 6PPD-Quinone, and IPPD

A structurally related subclass within this broader nitrogenous-organic family, the *para*-phenylenediamines (PPDs), has industrial applications that closely parallel those of the legacy contaminants described above, yet until recently received far less regulatory attention. Chemically, PPDs are closely related to diphenylamine and *N*-nitrosodiphenylamine. However, unlike diphenylamine, which features a single nitrogen atom linking two phenyl rings, the PPDs possess two nitrogen substituents positioned *para* to each other on a central benzene ring, with an additional alkyl or aryl group attached. PPDs are manufactured and used specifically as antiozonants and antioxidants for rubber, added to tires and other rubber products to prevent degradation from atmospheric ozone exposure (35). Two PPDs are of particular interest: *N*-(1,3-dimethylbutyl)-*N′*-phenyl-*p*-phenylenediamine (6PPD) and *N*-isopropyl-*N′*-phenyl-*p*-phenylenediamine (IPPD). Both serve as widely used tire-rubber antiozonants that migrate to the tire surface and react sacrificially with ozone to preserve the rubber matrix (36). As tires wear on roadways, 6PPD is shed in rubber particles that continue to react with ozone in the environment, forming the transformation product, 6PPD-quinone (6PPD-Q) (35,36).

These emerging compounds share core characteristics with the legacy SPL aromatic amines. Specifically, 6PPD, IPPD, and 6PPD-Q retain the same aromatic amine core structure, undergo related oxidative transformation pathways – culminating, in the case of 6PPD, in a reactive quinone-imine structure analogous to the toxic electrophilic intermediates of legacy aromatic

amines – and originate from the same industrial rubber and dye manufacturing sectors that generated historical aniline- and toluidine-contaminated NPL sites (28,35). Despite these shared properties and a rapidly growing body of toxicological evidence, neither 6PPD, 6PPD-Q, nor IPPD currently appears on the SPL, nor are they routinely reported as Superfund contaminants of concern; EPA's Office of Chemical Safety and Pollution Prevention has issued only non-binding aquatic-life screening values. However, in November 2024, EPA issued an advance notice of proposed rulemaking under the Toxic Substances Control Act (TSCA) to gather data on the environmental and health risks of 6PPD and 6PPD-Q (37). This regulatory gap, wherein chemicals structurally and mechanistically continuous with a well-established SPL contaminant class remain unlisted, defines 6PPD, 6PPD-Q, and IPPD as classic "*emerging contaminants*": harmful or potentially harmful substances actively detected in the environment but not yet formally regulated.

Nowhere is the significance of this emerging contaminant class clearer than in the Pacific Northwest. For over two decades, researchers observed a recurring, unexplained phenomenon in which adult coho salmon (*Oncorhynchus kisutch*) returning to spawn in urbanized Puget Sound streams died prematurely, with die-offs claiming up to 90% of returning runs within hours of stormwater exposure (38,39). This die-off was found to be strongly correlated with watershed imperviousness, road density, and traffic intensity, but its specific chemical cause eluded identification for years. In 2021, Tian et al. resolved this mystery, identifying 6PPD-Q as the primary causative agent. Through toxicity-directed fractionation of roadway runoff and non-target chemical analysis, 6PPD-Q, was determined to be acutely lethal to coho salmon at a median lethal concentration (LC50) of roughly 0.8 micrograms per liter – a concentration routinely measured and exceeded in urban Pacific Northwest streams (40).

### 2.4 Acquiring Superfund Data and an Agentic, Human-in-the-Loop Filtering Workflow

To construct this case study, we manually retrieved site- and contaminant-level records from EPA's Superfund Enterprise Management System (SEMS) – the database of record for Superfund site inventory, status, and contaminant information – using the public Superfund Site Information search tool hosted at cumulis.epa.gov (41). SEMS integrates data from EPA's legacy Superfund tracking systems and allowing queries and exports by site name, state, EPA region, NPL status, and contaminant. Current NPL sites for each region were queried and downloaded.

The exported dataset lists, for each NPL site, the identified contaminants along with the environmental media and exposure pathways for which each chemical is designated a contaminant of concern. Because this file contains hundreds of distinct substances across thousands of site-contaminant records, and because our research question required cross-referencing a multi-compound chemical class (multiple nitrogenous organic compounds) against human-health-specific designations and a specific four-state EPA region, we selected these data to demonstrate the HITL agentic AI framework. Rather than drafting the filtering code from scratch, we worked within an agentic LLM using Claude Desktop running Sonnet 5 Model (Anthropic) with local file access to the exported SEMS data. In natural language, we prompted the agent to identify NPL sites where one or more nitrogenous organic chemicals were listed as a contaminant of concern for human health and separately requested that it restrict results to EPA Region 10 (Alaska, Idaho, Oregon, and Washington). Although the agent rapidly returned site counts for the query, verifying and reproducing these results benefited from requesting standalone, executable R code. This step revealed key discrepancies that highlighted the necessity for human oversight. First, we observed a scope flaw: the generated code initially restricted filtering to Region 10 states, requiring manual script editing to extend the query to all NPL sites nationwide, so that both US-wide and Region 10-specific sites could be identified. Second, we observed a problem with semantic ambiguity: the AI applied its own broad interpretation of "*nitrogenous organic chemicals*," which deviated from our *a priori* target list. To eliminate chemical name ambiguity, we modified the script to filter strictly on Chemical Abstracts Service Registry Numbers (CAS RNs).

The final target list comprised a curated set of 15 CAS RNs. The revised R code was written flexibly to allow swapping of the CAS RN list, making the pipeline reusable for other Superfund contaminants of concern in the future. To verify the agent's interactive outputs, the code was executed outside the LLM environment in RStudio (v2026.05.1+225) using R (v4.5.2). Core code elements are detailed in the manuscript, with complete scripts provided in the Supplemental Information.

The verified analysis identified 181 NPL Superfund sites nationwide where target chemicals were listed as contaminants of concern for human health. Among these, 9 sites were located within EPA Region 10. Output datasets included site names, geographic locations (city and state), contaminated media (e.g., soil, groundwater), and direct links to EPA site summary pages to support follow-up community engagement and risk communication activities.

```
library(dplyr)

# US-wide.
filtered <- site_contams %>%
  filter(CASRN %in% target_cas,
         `Contaminant of Concern (Human Health)` == "Yes") %>%
  distinct(`Site Name.x`, `EPA ID`, `State.x`, CASRN, `Contaminant Name`, `Contaminant Media`,
.keep_all = TRUE) %>%
  select(`EPA ID`, `Site Name.x`, City, `State.x`,
         `Contaminant Name`, CASRN,
         `Contaminant Media`,
         `Contaminant of Concern (Human Health)`, `Contaminant of Concern (Ecological)`,
         `Site Type`,
         `Site Type Subcategory`,
         `Superfund Site Profile Page URL`)

length(unique(filtered$`EPA ID`))

# EPA Region 10 comprises Alaska, Idaho, Oregon, and Washington.
region10_filtered <- site_contams %>%
  filter(CASRN %in% target_cas,
         `Contaminant of Concern (Human Health)` == "Yes",
         Region == 10) %>%
  distinct(`Site Name.x`, `EPA ID`, `State.x`, CASRN, `Contaminant Name`, `Contaminant Media`,
.keep_all = TRUE) %>%
  select(`EPA ID`, `Site Name.x`, City, `State.x`,
```

```
         `Contaminant Name`, CASRN,
         `Contaminant Media`,
         `Contaminant of Concern (Human Health)`, `Contaminant of Concern (Ecological)`,
         `Site Type`,
         `Site Type Subcategory`,
         `Superfund Site Profile Page URL`)

length(unique(region10_filtered$`EPA ID`))
```

#### 2.5 Merging with ATSDR's Substance Priority List to Assess Population Exposure and Toxicity

To place the site-level results within a national toxicological and exposure context, we downloaded the 2025 Substance Priority List from the ATSDR website (26). As described previously, the SPL ranks substances detected at NPL sites using a composite score based on three factors: frequency of occurrence across NPL sites, toxicity, and potential for human exposure. The exposure component is further broken down into sub-factors reflecting the size of the population living near sites where the chemical is present.

Using the agentic LLM session, we prompted the agent to generate R code to filter the SPL dataset using our predefined set of 15 CAS RNs and requested that it utilize the flextable R package create a summary table and export it to a Word document file. However, we subsequently edited and executed generated R code manually to customize the exported columns, ensuring the output included the variables most relevant to assessing site counts, population exposure and toxicity (26) (Table 1).

```
library(dplyr)
library(flextable)

spl_of_interest <- spl_2025 %>%
  filter(CASRN %in% target_cas) %>%
  select(`Substance Name`, `CASRN`, `NPL Site Frequency`, `Toxicity Points`,
         `Population Near NPL Sites`, `Total Exposure Points`, `Total Points`) %>%
  arrange(-`Total Points`)

# Build and format the summary flextable
ft <- flextable(spl_of_interest) %>%
  theme_vanilla() %>%
  colformat_double(j = "Toxicity Points", digits = 2) %>%
  colformat_double(j = "Total Exposure Points", digits = 2) %>%
  colformat_double(j = "Total Points", digits = 2) %>%
  set_caption("Targeted NOCs at Superfund Sites with ATSDR SPL Toxicity and Exposure Scores")

# Explicitly set column widths (Inches)
ft <- width(ft, j = ~ `Substance Name`, width = 2.6)
ft <- width(ft, j = ~ CASRN, width = 0.8)

# Save the formatted table as a Word document
save_as_docx(ft, path = "noc_table_spl.docx")
```

**Table 1.** Selected Nitrogenous Organic Chemicals in the ATSDR Substance Priority List

| Substance Name | CASRN | NPL Site Frequency* | Toxicity Points** | Population Near NPL Sites** | Total Exposure Points** | Total Points** |
|---|---|---|---|---|---|---|
| CARBAZOLE | 86-74-8 | 96 | 404.54 | 2,018,757 | 424.94 | 1,229.58 |
| N-NITROSODI-N-PROPYLAMINE | 621-64-7 | 43 | 415.74 | 464,952 | 346.68 | 1,092.13 |
| N-NITROSODIPHENYLAMINE | 86-30-6 | 89 | 279.57 | 836,865 | 402.83 | 1,075.86 |
| NITROBENZENE | 98-95-3 | 57 | 342.41 | 389,013 | 377.03 | 1,073.85 |
| N-NITROSODIMETHYLAMINE | 62-75-9 | 18 | 452.97 | 1,013,768 | 321.20 | 1,027.54 |
| BENZIDINE | 92-87-5 | 16 | 481.57 | 199,227 | 301.29 | 1,025.90 |
| 3,3'-DICHLOROBENZIDINE | 91-94-1 | 27 | 364.30 | 380,767 | 322.48 | 975.69 |
| 4-NITROPHENOL | 100-02-7 | 38 | 286.26 | 337,085 | 323.87 | 928.99 |
| ANILINE | 62-53-3 | 25 | 282.41 | 305,721 | 330.51 | 895.08 |
| 1,2-DIPHENYLHYDRAZINE | 122-66-7 | 9 | 375.09 | 151,313 | 236.28 | 803.97 |
| 4-NITROANILINE | 100-01-6 | 10 | 305.94 | 69,254 | 245.60 | 753.37 |
| 4-AMINOBIPHENYL | 92-67-1 | 2 | 436.34 | 24,177 | 123.54 | 620.64 |
| DIPHENYLAMINE | 122-39-4 | 5 | 189.44 | 10,001 | 174.37 | 504.89 |

* ATSDR's Substance Priority List (SPL) includes the number of proposed or final National Priority List (NPL) sites at which the substance was found.

** ATSDR's SPL computes Toxicity Points based on cancer and non-cancer inhalation and oral toxicity weights, Population Near NPL Sites based on a 1-mile radius, Total Exposure Points based on a Complete Exposure Pathway (CEP) analysis, and Total Points based on a composite calculation of Site Frequency Points, Toxicity Points, and Total Exposure Points (For comparison the #1 Ranked SPL (Lead) has a Total Point value of 1795.62).

### 2.6 Geospatial AI Mapping of Sites with Nitrogenous Organic Chemicals

To visualize the spatial distribution of the identified Superfund sites across the U.S., where nitrogenous organic chemicals have been detected, we created a point-location map overlaid on state boundaries. The underlying spatial data were downloaded *a priori* from the EPA's Facility Registry Service (FRS) as a geodatabase (42). Because the FRS geodatabase contains multiple spatial layers, we used the sf package in R (43) to inspect the layer contents and extract the SEMS_NPL layer, which contains spatial point geometries and facility identifiers for NPL sites.

After exporting the SEMS_NPL layer, we used the agentic AI session to prompt the agent to generate R code that would plot the filtered NPL sites over U.S. state boundaries. When executed in RStudio, the generated code successfully rendered the desired map. Through a subsequent prompt, we instructed the agent to modify the code to export and save the final map as a publication-ready PDF file (Figure 2).

```
library(sf)
library(dplyr)
library(ggplot2)


# US state boundaries
us_states <- st_as_sf(maps::map("state", plot = FALSE, fill = TRUE))


# Load the SEMS_NPL point geometries and the US-wide filtered sites
sems <- readRDS(file = "sems_sf.rds")

site_summary <- read.csv(file = "US_nitrogenous_organics_site_summary.csv",
                          stringsAsFactors = FALSE)

coc_sites <- sems %>%
  inner_join(site_summary, by = c("PGM_SYS_ID" = "EPA.ID"))

# align CRS so the points and state polygons overlay correctly
coc_sites <- st_transform(coc_sites, st_crs(us_states))


# Plot filtered sites on top of state boundaries
test_map <- ggplot() +
  geom_sf(data = us_states, fill = "gray95", color = "white") +
  geom_sf(data = coc_sites, color = "firebrick", size = 1, alpha = 0.7) +
  coord_sf(xlim = c(-125, -66), ylim = c(24, 50)) +  # crop to CONUS
  theme_minimal() +
  labs(title = "US NPL Sites with Nitrogenous Organic Contaminants of Concern",
       x = NULL, y = NULL)

test_map

# save to PDF file
ggsave(filename = "test_map.pdf", plot = test_map, width = 11, height = 8.5, units = "in")
```

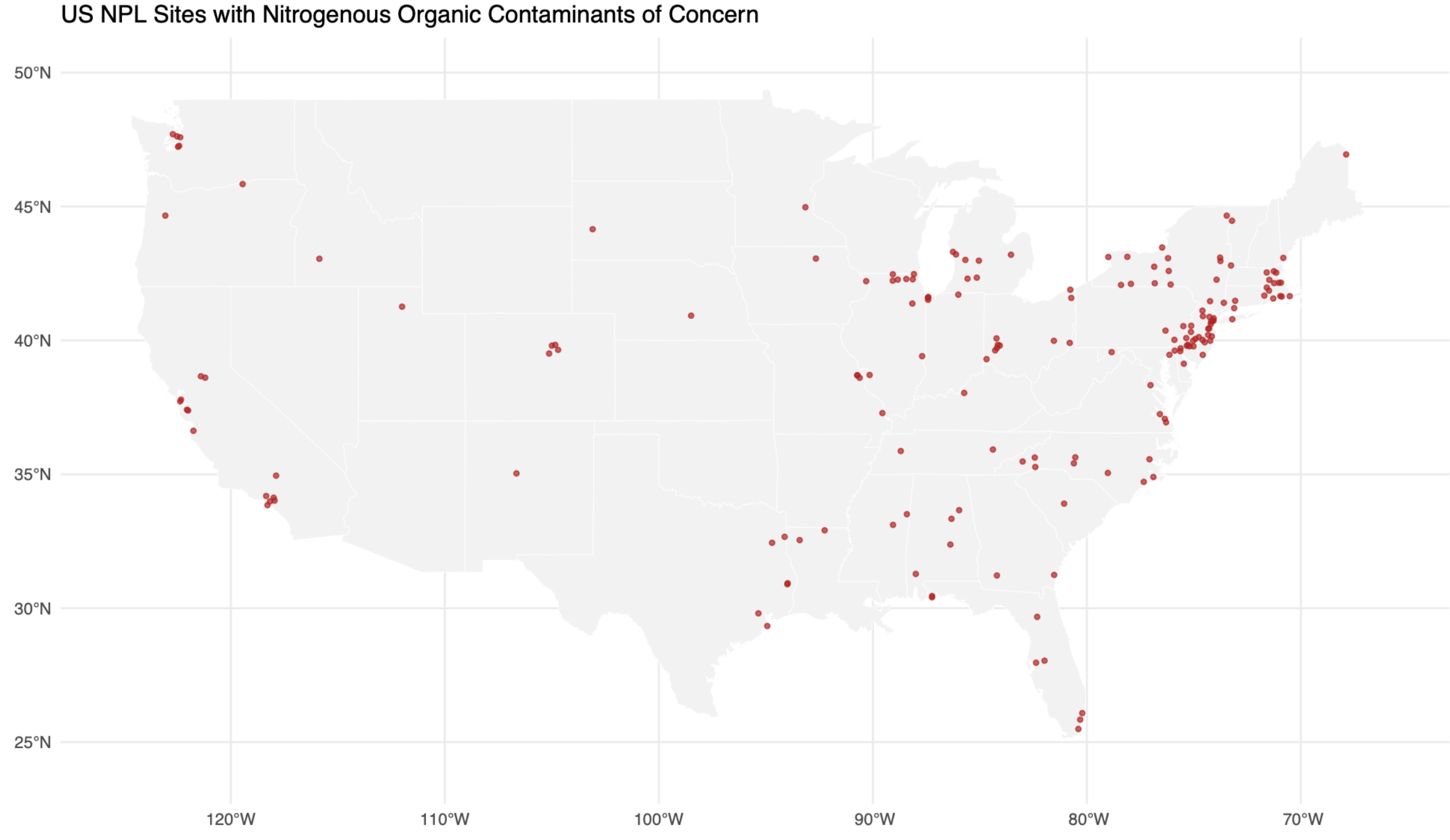

**Figure 2.** National Priority Sites with Nitrogenous Organic Contaminants of Concern for Human Health.

Separate from this static map, we asked the agent to generate code for an interactive map using the tmap R package (44) in its interactive view mode, enabling detailed exploration of individual site locations. In this instance, the generated code initially failed to execute because the AI agent incorporated deprecated tmap function parameters. After manually updating the syntax, the interactive map was successfully generated. Both the static and interactive maps illustrated that this contaminant class is not confined to the Pacific Northwest, but occurs at NPL sites nationwide.

```
library(tmap)

tmap_mode("view") # interactive
tm_shape(coc_sites) + tm_dots()
```

Executing more complex geospatial AI workflows, such as intersecting census tract-level population estimates from the American Community Survey with 3-mile buffer zones surrounding each Superfund site, proved feasible via AI-generated R code, though execution required a valid Census Application Programming Interface (API) key.

### 2.7 Cluster Analysis of Nitrogenous Organic Chemical Mixtures

For the fourth agentic AI analysis, we performed a multi-pollutant cluster analysis to evaluate co-occurring mixtures of nitrogenous organic contaminants. Given the common industrial origins, we anticipated that these compounds would frequently co-occur at Superfund sites. Real-world contamination is rarely limited to a single chemical; people and ecosystems are routinely exposed to complex and correlated chemical mixtures, and cluster analysis serves as a standard tool for identifying co-occurrence patterns (45,46). We prompted the agentic LLM to perform a cluster analysis of the nationwide filtered dataset established in previous steps.

The initial attempt clearly demonstrated the necessity of the HITL checkpoint. The agent executed the cluster analysis directly on the SEMS export in its original form, which contains one row per site-chemical-media combination (e.g., a single site with benzidine detected in both soil and groundwater contributes two rows, not one). Clustering directly on this long-format table treated each site-chemical-media record as an independent observation rather than representing each site as a single observation characterized by its chemical profile. Consequently, the resulting clusters failed to represent coherent site groupings. We identified this error when the total count of clustered sites did not match the 181 sites established during the filtering phase. The AI agent did not detect this discrepancy and proceeded to complete the analysis, including generating summary figures from the flawed output.

To correct this issue, we instructed the agent to restructure the long SEMS dataset into a wide, one-row-per-site format, adhering to established data-tidying principles for long-to-wide transformations (47). Each of the 15 target chemicals was reshaped into a binary column – coded 1 if present anywhere at the site, and 0 otherwise – thereby collapsing across media types so that each site's chemical profile defined the observation unit. Using this wide binary matrix, we prompted the agent to calculate Jaccard distance matrices (appropriate for binary presence-absence data) and perform hierarchical clustering, an approach well-supported in chemical mixture and co-exposure literature (45,46). We reviewed and executed this revised R code, verifying that the cluster sizes summed correctly to the 181 unique NPL sites across the U.S., confirming that the clustering operated on the intended unit of analysis.

The corrected results (Figure 3) revealed relatively few sites where only a single nitrogenous organic chemical was listed as a contaminant of concern. Instead, most sites grouped into a small number of distinct clusters characterized by multi-chemical mixtures. Following manual inspection of each cluster's composition, we assigned descriptive semantic labels to each group and prompted the AI agent to update the R script to generate a correspondingly labeled heatmap (Figure 3). This pattern confirmed our expectation that compounds sharing industrial lineages in dye, rubber, and chemical manufacturing frequently co-occur. Notably, three primary mixture clusters accounted for 76.8% (139 of 181) of the sites: the *N*-Nitrosodiphenylamine-mix, Carbazole-mix, and Nitrobenzene-mix clusters.

```r
library(dplyr)
library(tidyr)
library(proxy)
library(ggplot2)
library(scales)

# Load the US-wide filtered long-format export (see NOC_filtering.R)
filtered <- read.csv(file = "US_nitrogenous_organics_npl_filtered.csv",
                     stringsAsFactors = FALSE)

# Reshape the long export to one row per site (wide, binary)
# Collapse across contaminated media so a site's chemical profile, not its
# number of underlying rows, defines the observation used for clustering.
site_chemical_wide <- filtered %>%
  distinct(EPA.ID, Site.Name.x, State.x, CASRN) %>%
  mutate(present = 1L) %>%
  pivot_wider(
    id_cols = c(EPA.ID, Site.Name.x, State.x),
    names_from = CASRN,
    values_from = present,
    values_fill = 0L
  )

# Binary presence/absence matrix for clustering (0/1)
chem_matrix <- as.matrix(site_chemical_wide[, target_cas])

# Yes/No labeled wide table for review/output
site_chemical_wide_labels <- site_chemical_wide %>%
  mutate(across(all_of(target_cas), ~ ifelse(. == 1L, "Yes", "No"))) %>%
  rename_with(~ cas_names[.x], all_of(target_cas))

# Compute Jaccard distance on the binary chemical profile
jaccard_dist <- proxy::dist(chem_matrix, method = "Jaccard")

# Hierarchical clustering
hc <- hclust(jaccard_dist, method = "average")

# inspect the dendrogram to choose k before finalizing
plot(hc, labels = FALSE, main = "Dendrogram of NPL Site Chemical Profiles (Jaccard distance)")

k <- 10   # number of clusters; adjust after inspecting the dendrogram above
site_chemical_wide$cluster <- cutree(hc, k = k)

# Compute per-cluster chemical prevalence (proportion of sites)
cluster_prevalence <- site_chemical_wide %>%
  group_by(cluster) %>%
  summarize(across(all_of(target_cas), mean), n_sites = n(), .groups = "drop") %>%
  mutate(cluster_label = paste0("Cluster ", cluster, " (n=", n_sites, ")")) %>%
  select(cluster_label, all_of(target_cas)) %>%
  rename_with(~ cas_names[.x], all_of(target_cas))
```

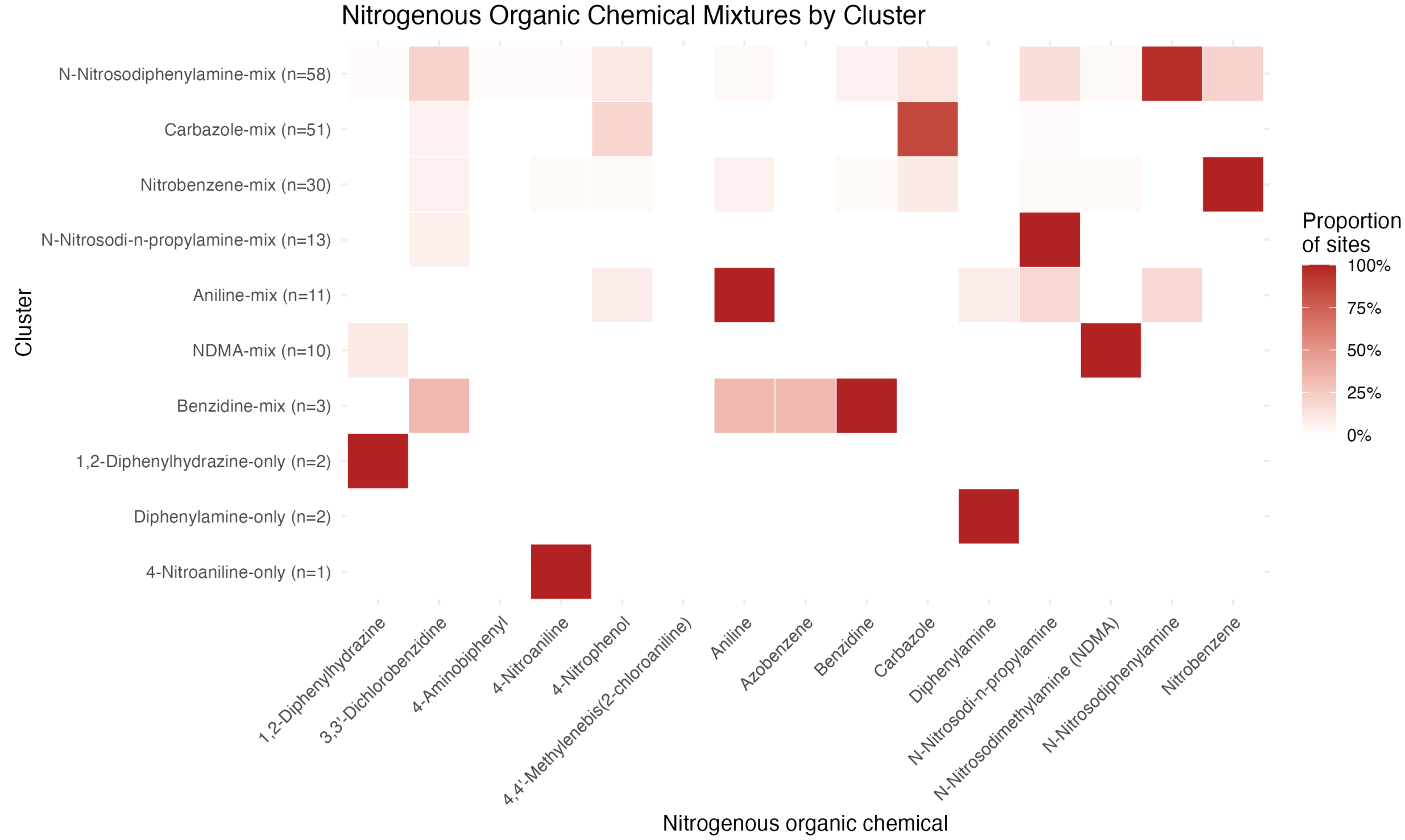

**Figure 3.** Heat map showing proportion of Superfund sites in each cluster (y-axis) that include specific nitrogenous organic contaminants (x-axis).

## 3. Lessons Learned and Future Opportunities

This case study demonstrates that agentic AI, when integrated with real-world environmental health data and tools, is fully capable of executing practical, multi-step analytical workflows representative of every-day research: wrangling and filtering large administrative datasets, merging primary data with independent secondary sources for contextualization, performing geospatial analyses and mapping, and identifying multi-dimensional co-occurrence patterns via cluster analysis.

However, this case study just as clearly highlights the limitations of taking agentic AI outputs at face value. Across the different workflow stages – most notably during the cluster analysis, but also in subtle ways throughout earlier steps – the code and outputs generated by the agent contained critical errors that were not obvious from the model's self-generated narrative descriptions. Identifying these errors, and either prompting the agent to correct them or manually refactoring the code, depended entirely on active HITL oversight. Left unchecked, the agent's confident narrative summaries would have masked underlying analytical flaws. This finding aligns with a growing body of empirical literature demonstrating that current LLMs exhibit lower accuracy on realistic, multi-step data science coding tasks than their conversational fluency implies, as well as classic research on automation bias showing that human overseers fail to catch errors unless they remain critically engaged with automated outputs rather than passively accepting them (48,49).

Furthermore, our experience underscores a critical, often under-appreciated prerequisite of the agentic AI framework: the human participant must contribute genuine domain-specific expertise, rather than serving merely as a generic validator of code syntax. In this case study, framing an appropriate sequence of research questions, such as defining a chemically and toxicologically coherent set of nitrogenous organic contaminants, navigating the regulatory distinctions between the NPL and the SPL, and conceptualizing environmental contamination as a multi-pollutant mixture problem, required domain knowledge that was provided by the human researchers rather than supplied by the agent.

Crucially, successful oversight required a granular understanding of the underlying SEMS data structure. Recognizing that the SEMS export contained one row per site-chemical-medium combination enabled the human researcher to diagnose why the agent's initial long-format cluster analysis was flawed. Effective implementation also depended on technical literacy: the ability to specify appropriate R packages, functions, and statistical methods in prompts, to critically audit generated code rather than relying on natural-language summaries, and to identify when syntactically valid code produced methodologically invalid results. Consistent with recent research on prompt engineering for scientific applications, the utility of agentic AI collaborations in environmental health appears to rely less on the raw capability of the underlying LLM and more on the domain expertise, data familiarity, and technical fluency of the human directing it (50). In this sense, directing an agentic AI system closely mirrors effective interdisciplinary communication within traditional team science workflows.

Several opportunities follow from these lessons for strengthening this framework going forward. First, structured, checklist-based verification protocols – an agentic analogue of the code review practices standard in software engineering and scientific computing – could formalize what was, in this case study, an ad hoc process of the human researcher identifying errors. This would help extend the framework's benefits to investigators who are still developing domain or technical fluency. Second, equipping AI agents with direct, tool-based access to authoritative data sources could reduce reliance on manual, one-off file downloads while preserving the HITL checkpoints that ensure data integrity. Finally, because this framework explicitly mirrors the traditional mentorship structure of a research team, it suggests a natural role for agentic AI in training future environmental health researchers. Rather than replacing mentorship between senior and junior investigators, agentic AI can serve as an additional collaborator whose outputs junior researchers practice reviewing critically under senior supervision.

To explore how this framework might extend beyond our initial case study, we asked the Claude AI agent – after providing full context from all previous analytical steps – to propose additional research directions. The agent suggested several compelling avenues. It proposed an environmental justice analysis with EJScreen or the Social Vulnerability Index to test whether sites with nitrogenous organic contaminants of concern disproportionately burden low-income or minority communities. It also suggested a temporal trend analysis of NPL listing and remediation dates to connect clusters of site listings to historical periods of dye, rubber, and chemical manufacturing activity. It recommended a cross-class mixture analysis to extend the clustering approach beyond the nitrogenous organic chemicals to the full range of contaminant classes on the SPL, producing a fuller cumulative-risk picture of site-level contamination than any single chemical family can offer on its own. It also recommended site-level population exposure modeling, using spatial buffers around each site and intersecting them with Census population data to move beyond the SPL's national aggregate exposure score toward a site-specific exposure estimate. And, drawing on a different capability of the underlying language model entirely, it suggested an analysis using natural-language extraction to mine the narrative text of Superfund Records of Decision to surface facility history, remedy rationale, and exposure pathway detail that are missing from structured databases like SEMS.

While proposing these additional analyses does not imply that AI possesses genuine scientific creativity, each suggestion is methodologically sound, feasible, and aligned with practical Superfund stakeholder concerns. Asking an AI collaborator not merely to execute an analysis, but to assist in conceptualizing the next one, highlights a broader opportunity: agentic AI, applied with critical, expertise-grounded human oversight, can substantially expand the analytical scope and efficiency of environmental health researchers. That prospect represents the most encouraging opportunity this framework offers to the field.

## Acknowledgements

We thank the CDC and US EPA for providing openly accessible data that were used for the case study.

## Funding

The authors did not receive any funding for this work.

**Supplemental Information**

# Agentic Artificial Intelligence for Reproducible Human-in-the-Loop Environmental Health Research

Edmund Seto

| | |
|---|---|
| Combine_cumulis_files.R | Combines the downloaded SEMS exports from the EPA Cumulis website query. |
| Read_ATSDR.R | Reads the ATSDR Excel file from the SPL website. |
| Read_FRS_geodatabase.R | Reads the EPA FRS geodatabase file. |
| NOC_filtering.R | Filters the NPL sites for NOCs of interest. |
| NOC_table.R | Creates summary table of the NOCs of interest from the SPL. |
| NOC_map.R and test_map.R | Two versions of the mapping script (one including interactive map). |
| NOC_cluster_analysis.R | Multi-chemical mixtures cluster analysis. |

## 1. Combine_cumulis_files.R

```r
##############################
# Combine_cumulis_files.R
#
# Edmund Seto
# 2026-07-30
##############################


# EPA cumulis data were downloaded from
https://cumulis.epa.gov/supercpad/CurSites/srchsites.cfm
#   for "NPL or Superfund Alternative Approach (SAA) Status:*" = Currently on the NPL
#   had to download each region separately.

# downloads are Excel files.  First sheet is site info.  Second sheet is contaminant info

library(readxl)
library(dplyr)

Region_1 <- read_excel("cumulis/Region_1.xls",
                       sheet = "Site Results", skip = 6)

Region_1c <- read_excel("cumulis/Region_1.xls",
                       sheet = "Contaminant Results_1", skip = 6)

Region_2 <- read_excel("cumulis/Region_2.xls",
                       sheet = "Site Results", skip = 6)

Region_2c <- read_excel("cumulis/Region_2.xls",
                        sheet = "Contaminant Results_1", skip = 6)

Region_3 <- read_excel("cumulis/Region_3.xls",
                       sheet = "Site Results", skip = 6)

Region_3c <- read_excel("cumulis/Region_3.xls",
                        sheet = "Contaminant Results_1", skip = 6)

Region_4 <- read_excel("cumulis/Region_4.xls",
                       sheet = "Site Results", skip = 6)

Region_4c <- read_excel("cumulis/Region_4.xls",
                        sheet = "Contaminant Results_1", skip = 6)

Region_5 <- read_excel("cumulis/Region_5.xls",
                       sheet = "Site Results", skip = 6)

Region_5c <- read_excel("cumulis/Region_5.xls",
                        sheet = "Contaminant Results_1", skip = 6)

Region_6 <- read_excel("cumulis/Region_6.xls",
                       sheet = "Site Results", skip = 6)

Region_6c <- read_excel("cumulis/Region_6.xls",
```

```r
                         sheet = "Contaminant Results_1", skip = 6)

Region_7 <- read_excel("cumulis/Region_7.xls",
                        sheet = "Site Results", skip = 6)

Region_7c <- read_excel("cumulis/Region_7.xls",
                         sheet = "Contaminant Results_1", skip = 6)

Region_8 <- read_excel("cumulis/Region_8.xls",
                        sheet = "Site Results", skip = 6)

Region_8c <- read_excel("cumulis/Region_8.xls",
                         sheet = "Contaminant Results_1", skip = 6)

Region_9 <- read_excel("cumulis/Region_9.xls",
                        sheet = "Site Results", skip = 6)

Region_9c <- read_excel("cumulis/Region_9.xls",
                         sheet = "Contaminant Results_1", skip = 6)

Region_10 <- read_excel("cumulis/Region_10.xls",
                        sheet = "Site Results", skip = 6)

Region_10c <- read_excel("cumulis/Region_10.xls",
                         sheet = "Contaminant Results_1", skip = 6)

# bind rows

Region_sites <- bind_rows(Region_1, Region_2, Region_3, Region_4, Region_5, Region_6,
Region_7, Region_8, Region_9, Region_10)
Region_site_contams <- bind_rows(Region_1c, Region_2c, Region_3c, Region_4c, Region_5c,
Region_6c, Region_7c, Region_8c, Region_9c, Region_10c)


# merged contaminants_site with site info
site_contams <- Region_site_contams %>%
  left_join(Region_sites, by="EPA ID")

# save to file
saveRDS(site_contams, file = "site_contams.rds")
```

### 3. Read_ATSDR.R

```r
##########################
# Read_ATSDR.R
#
# Edmund Seto
# 2026-07-30
##########################


# reads the ATSDR's Substance priority list (SPL) Excel file
#  https://www.atsdr.cdc.gov/programs/substance-priority-list.html

library(readxl)

ATSDR_2025_Official_SPL <- read_excel("ATSDR/ATSDR-2025-Official-SPL.xlsx",
                                      sheet = "SPL Data")

# save as RDS
saveRDS(ATSDR_2025_Official_SPL, file="spl_2025.rds")
```

## 5. Read_FRS_geodatabase.R

```r
################################
# Read_FRS_geodatabase.R
#
# Edmund Seto
# 2026-07-30
################################

# reads the EPA FRS geodatabase downloaded from data.gov
#   https://catalog.data.gov/dataset/epa-facility-registry-service-frs-sems

library(sf)

gdb_path <- "./data.gov/FRS_INTERESTS.gdb"

# what's in it?
st_layers(gdb_path)

# focus on the SEMS_NPL

# Option A: Standard read (prints a summary of the data layout in the console)
sems <- st_read(dsn = gdb_path, layer = "SEMS_NPL")

# Option B: Silent read (returns data cleanly as a spatial tibble dataframe)
# my_layer <- read_sf(dsn = gdb_path, layer = "your_layer_name")

saveRDS(sems, file="sems_sf.rds")
```

## 2. NOC_filtering.R

```r
##########################
# NOC_filtering.R
#
# Edmund Seto
# 2026-08-05
##########################

# Filter SEMS/cumulis export (see Combine_cumulis_files.R) to nitrogenous
# organic contaminants of concern (human health) at Final NPL sites in
# US and EPA Region 10 (Alaska, Idaho, Oregon, Washington).
#
# Implements the filtering workflow described in Section 2.5 of the paper
# ("Methods: Acquiring SEMS Data and an Agentic, Human-in-the-Loop Filtering
# Workflow"). Column names below reflect the structure of the combined
# cumulis/SEMS export produced by Combine_cumulis_files.R.

library(dplyr)


# --- 1. Load the combined site-contaminant data ---
# Produced by Combine_cumulis_files.R from the manually exported SEMS/cumulis
# regional Excel files (site info + contaminant info, joined on EPA ID).
# All records reflect sites "Currently on the NPL" (Final NPL) per the
# cumulis search criteria used at export time (see Combine_cumulis_files.R).
site_contams <- readRDS(file = "site_contams.rds")


# --- 2. Define the target list of nitrogenous organic chemicals by CAS number ---
# Curated target list spanning five related nitrogen-bearing chemical
# subclasses (Section 2.2): aromatic amines, hydrazines/azo compounds,
# nitroaromatics, an N-heterocyclic aromatic, and N-nitrosamines.
target_cas <- c(
  "92-87-5",    # Benzidine (aromatic amine)
  "62-53-3",    # Aniline (aromatic amine)
  "91-94-1",    # 3,3'-Dichlorobenzidine (aromatic amine)
  "92-67-1",    # 4-Aminobiphenyl (aromatic amine)
  "101-14-4",   # 4,4'-Methylenebis(2-chloroaniline) / MBOCA (aromatic amine)
  "122-39-4",   # Diphenylamine (aromatic amine, secondary)
  "122-66-7",   # 1,2-Diphenylhydrazine / hydrazobenzene (hydrazine)
  "103-33-3",   # Azobenzene (azo compound)
  "98-95-3",    # Nitrobenzene (nitroaromatic)
  "100-01-6",   # 4-Nitroaniline (nitroaromatic)
  "100-02-7",   # 4-Nitrophenol (nitroaromatic)
  "86-74-8",    # Carbazole (N-heterocyclic aromatic)
  "86-30-6",    # N-Nitrosodiphenylamine (N-nitrosamine)
  "62-75-9",    # N-Nitrosodimethylamine / NDMA (N-nitrosamine)
  "621-64-7"    # N-Nitrosodi-n-propylamine (N-nitrosamine)
)


# --- 3. Filter to target CAS numbers, human-health COC ---
```

```
# US-wide.
filtered <- site_contams %>%
  filter(CASRN %in% target_cas,
         `Contaminant of Concern (Human Health)` == "Yes") %>%
  distinct(`Site Name.x`, `EPA ID`, `State.x`, CASRN, `Contaminant Name`, `Contaminant
Media`, .keep_all = TRUE) %>%
  select(`EPA ID`, `Site Name.x`, City, `State.x`,
         `Contaminant Name`, CASRN,
         `Contaminant Media`,
         `Contaminant of Concern (Human Health)`, `Contaminant of Concern (Ecological)`,
         `Site Type`,
         `Site Type Subcategory`,
         `Superfund Site Profile Page URL`)

length(unique(filtered$`EPA ID`))

# EPA Region 10 comprises Alaska, Idaho, Oregon, and Washington.
region10_filtered <- site_contams %>%
  filter(CASRN %in% target_cas,
         `Contaminant of Concern (Human Health)` == "Yes",
         Region == 10) %>%
  distinct(`Site Name.x`, `EPA ID`, `State.x`, CASRN, `Contaminant Name`, `Contaminant
Media`, .keep_all = TRUE) %>%
  select(`EPA ID`, `Site Name.x`, City, `State.x`,
         `Contaminant Name`, CASRN,
         `Contaminant Media`,
         `Contaminant of Concern (Human Health)`, `Contaminant of Concern (Ecological)`,
         `Site Type`,
         `Site Type Subcategory`,
         `Superfund Site Profile Page URL`)

length(unique(region10_filtered$`EPA ID`))


# --- 4. Summarize sites and chemical frequency for review ---

# US-wide.
site_summary <- filtered %>%
  group_by(`Site Name.x`, `EPA ID`, `State.x`) %>%
  summarize(chemicals_detected = paste(sort(unique(`Contaminant Name`)), collapse = ";
"),
            n_chemicals = n_distinct(CASRN),
            .groups = "drop")

write.csv(filtered, file = "US_nitrogenous_organics_npl_filtered.csv", row.names = FALSE)
write.csv(site_summary, file = "US_nitrogenous_organics_site_summary.csv", row.names =
FALSE)

# Region 10-specific.
site_summary_region10 <- region10_filtered %>%
  group_by(`Site Name.x`, `EPA ID`, `State.x`) %>%
  summarize(chemicals_detected = paste(sort(unique(`Contaminant Name`)), collapse = ";
"),
```

```
            n_chemicals = n_distinct(CASRN),
            .groups = "drop")

write.csv(region10_filtered, file = "region10_nitrogenous_organics_npl_filtered.csv",
row.names = FALSE)
write.csv(site_summary_region10, file = "region10_nitrogenous_organics_site_summary.csv",
row.names = FALSE)
```

### 4. NOC_table.R

```
##########################
# NOC_table.R
#
# Edmund Seto
# 2026-08-05
##########################

# Reads the ATSDR's Substance Priority List
# (Section 2.6) and filters by CAS number to build a formatted summary table with
# flextable.
#
# Per Section 2.6, the SPL's "NPL Site Frequency" counts occurrence across
# both final and proposed NPL sites nationwide.

library(dplyr)
library(flextable)


# --- 1. Load the ATSDR SPL data (see Read_ATSDR.R) ---
# Same 15 target CAS numbers used in Section 2.5 / NOC_filtering.R.
spl_2025 <- readRDS(file = "spl_2025.rds")

target_cas <- c(
  "92-87-5",    # Benzidine (aromatic amine)
  "62-53-3",    # Aniline (aromatic amine)
  "91-94-1",    # 3,3'-Dichlorobenzidine (aromatic amine)
  "92-67-1",    # 4-Aminobiphenyl (aromatic amine)
  "101-14-4",   # 4,4'-Methylenebis(2-chloroaniline) / MBOCA (aromatic amine)
  "122-39-4",   # Diphenylamine (aromatic amine, secondary)
  "122-66-7",   # 1,2-Diphenylhydrazine / hydrazobenzene (hydrazine)
  "103-33-3",   # Azobenzene (azo compound)
  "98-95-3",    # Nitrobenzene (nitroaromatic)
  "100-01-6",   # 4-Nitroaniline (nitroaromatic)
  "100-02-7",   # 4-Nitrophenol (nitroaromatic)
  "86-74-8",    # Carbazole (N-heterocyclic aromatic)
  "86-30-6",    # N-Nitrosodiphenylamine (N-nitrosamine)
  "62-75-9",    # N-Nitrosodimethylamine / NDMA (N-nitrosamine)
  "621-64-7"    # N-Nitrosodi-n-propylamine (N-nitrosamine)
)

spl_of_interest <- spl_2025 %>%
  filter(CASRN %in% target_cas) %>%
  select(`Substance Name`, `CASRN`, `NPL Site Frequency`, `Toxicity Points`,
         `Population Near NPL Sites`, `Total Exposure Points`, `Total Points`) %>%
  arrange(-`Total Points`)


write.csv(spl_of_interest, file = "US_nitrogenous_organics_spl.csv", row.names = FALSE)

# --- 2. Build and format the summary flextable ---
ft <- flextable(spl_of_interest) %>%
  theme_vanilla() %>%
```

```
  colformat_double(j = "Toxicity Points", digits = 2) %>%
  colformat_double(j = "Total Exposure Points", digits = 2) %>%
  colformat_double(j = "Total Points", digits = 2) %>%
  set_caption("Targeted NOCs at Superfund Sites with ATSDR SPL Toxicity and Exposure
Scores")

ft

# Explicitly set column widths (Inches)
ft <- width(ft, j = ~ `Substance Name`, width = 2.6)
ft <- width(ft, j = ~ CASRN, width = 0.8)
ft

# Save the formatted table as a Word document
save_as_docx(ft, path = "noc_table_spl.docx")
```

### 6. NOC_map.R

```r
##########################
# NOC_map.R
#
# Edmund Seto
# 2026-08-05
##########################

# Build interactive and static maps of NPL sites nationwide where one or
# more of the 15 target nitrogenous organic chemicals is listed as a
# contaminant of concern for human health (Section 2.7), using the same
# tmap + usmap mapping approach as Search_by_contam.R.

library(sf)
library(dplyr)
library(tmap)
library(usmap)


# --- 1. Load the SEMS_NPL point geometries from the FRS geodatabase ---
# Extracted in Read_FRS_geodatabase.R (st_layers() confirms available layers;
# SEMS_NPL provides point-location geometry and facility identifiers for
# sites on the Superfund National Priorities List).
sems <- readRDS(file = "sems_sf.rds")
str(sems)


# --- 2. Load the nationwide filtered site summary (see NOC_filtering.R) ---
# Same 15 target CAS numbers, human-health COC flag, and Final NPL status
# as Section 2.5, applied across all EPA regions (no Region 10 restriction)
# so the resulting site list spans the entire country.
site_summary <- read.csv(file = "US_nitrogenous_organics_site_summary.csv",
                          stringsAsFactors = FALSE)


# --- 3. Join the filtered site list to SEMS_NPL point geometries ---
# Join key reflects the FRS geodatabase schema (PGM_SYS_ID) against the
# EPA ID field carried through the cumulis/SEMS export.
coc_sites <- sems %>%
  right_join(site_summary, join_by("PGM_SYS_ID" == "EPA.ID"))

str(coc_sites)    # it's still sf
st_crs(coc_sites) # ID["EPSG",4269]] # NAD83

# note 2 missing sites due to no match with FRS database (one in NJ, one in ID)


# --- 4. Interactive map for exploring site locations ---
tmap_mode("view") # interactive

tm_shape(coc_sites) + tm_dots()
```

```r
# --- 5. Static map for presentations and publications ---
tmap_mode("plot") # static

us_map_data <- us_map()
# Convert the data frame to a spatial 'sf' object, which tmap can use directly
# us_map() data is already projected correctly by the package
us_sf <- st_as_sf(us_map_data, coords = c("x", "y"), crs = usmap_crs())

us_sf_conus <- us_sf %>%
  filter(!abbr %in% c("AK", "HI", "PR"))

my_map <- tm_shape(us_sf_conus) +
  tm_polygons(col = "lightgray", border.col = "white") +
  tm_shape(coc_sites) +
  tm_dots()

my_map

# export
tmap_save(my_map, filename = "map_output_us.pdf", width = 11, height = 8.5, units = "in")
```

### 7. test_map.R

```r
##########################
# test_map.R
#
# Edmund Seto
# 2026-08-05
##########################

# Quick test map: plots the US-wide filtered NPL sites (see NOC_filtering.R)
# on top of US state boundaries, using ggplot2 + sf.

library(sf)
library(dplyr)
library(ggplot2)


# --- 1. US state boundaries ---
# Base R's "maps" package ships state polygons with no external download or
# API key required; convert to sf so it can be plotted with geom_sf().
us_states <- st_as_sf(maps::map("state", plot = FALSE, fill = TRUE))


# --- 2. Load the SEMS_NPL point geometries and the US-wide filtered sites ---
sems <- readRDS(file = "sems_sf.rds")

site_summary <- read.csv(file = "US_nitrogenous_organics_site_summary.csv",
                         stringsAsFactors = FALSE)

coc_sites <- sems %>%
  inner_join(site_summary, by = c("PGM_SYS_ID" = "EPA.ID"))

# align CRS so the points and state polygons overlay correctly
coc_sites <- st_transform(coc_sites, st_crs(us_states))


# --- 3. Plot filtered sites on top of state boundaries ---
test_map <- ggplot() +
  geom_sf(data = us_states, fill = "gray95", color = "white") +
  geom_sf(data = coc_sites, color = "firebrick", size = 1, alpha = 0.7) +
  coord_sf(xlim = c(-125, -66), ylim = c(24, 50)) +  # crop to CONUS
  theme_minimal() +
  labs(title = "US NPL Sites with Nitrogenous Organic Contaminants of Concern",
       x = NULL, y = NULL)

test_map

ggsave(filename = "test_map.pdf", plot = test_map, width = 11, height = 8.5, units =
"in")
```

### 8. NOC_cluster_analysis.R

```
##########################
# NOC_cluster_analysis.R
#
# Edmund Seto
# 2026-08-05
##########################

# Cluster analysis of nitrogenous organic chemical co-occurrence at NPL
# sites nationwide (Section 2.8). Reshapes the long, site-chemical-media
# US-wide filtered export (see NOC_filtering.R) into a wide, one-row-per-site
# binary presence/absence profile across the 15 target CAS chemicals, then
# computes Jaccard distance and performs hierarchical clustering on that
# site-level chemical profile -- not on the long-format export directly,
# which would fragment each site across its underlying chemical-media rows.

library(dplyr)
library(tidyr)
library(proxy)
library(ggplot2)
library(scales)


# --- 1. Load the US-wide filtered long-format export (see NOC_filtering.R) ---
filtered <- read.csv(file = "US_nitrogenous_organics_npl_filtered.csv",
                      stringsAsFactors = FALSE)


# --- 2. Target CAS numbers and readable chemical names ---
# Same 15 target CAS numbers used throughout Sections 2.5-2.7.
cas_names <- c(
  "92-87-5"  = "Benzidine",
  "62-53-3"  = "Aniline",
  "91-94-1"  = "3,3'-Dichlorobenzidine",
  "92-67-1"  = "4-Aminobiphenyl",
  "101-14-4" = "4,4'-Methylenebis(2-chloroaniline)",
  "122-39-4" = "Diphenylamine",
  "122-66-7" = "1,2-Diphenylhydrazine",
  "103-33-3" = "Azobenzene",
  "98-95-3"  = "Nitrobenzene",
  "100-01-6" = "4-Nitroaniline",
  "100-02-7" = "4-Nitrophenol",
  "86-74-8"  = "Carbazole",
  "86-30-6"  = "N-Nitrosodiphenylamine",
  "62-75-9"  = "N-Nitrosodimethylamine (NDMA)",
  "621-64-7" = "N-Nitrosodi-n-propylamine"
)
target_cas <- names(cas_names)


# --- 3. Reshape the long export to one row per site (wide, binary) ---
# Collapse across contaminated media so a site's chemical profile, not its
# number of underlying rows, defines the observation used for clustering.
```

```r
site_chemical_wide <- filtered %>%
  distinct(EPA.ID, Site.Name.x, State.x, CASRN) %>%
  mutate(present = 1L) %>%
  pivot_wider(
    id_cols = c(EPA.ID, Site.Name.x, State.x),
    names_from = CASRN,
    values_from = present,
    values_fill = 0L
  )

# ensure all 15 target CAS columns exist, even if a chemical was present at
# zero sites nationwide
missing_cas <- setdiff(target_cas, names(site_chemical_wide))
for (cas in missing_cas) site_chemical_wide[[cas]] <- 0L


# --- 4. Sanity check: one row per unique site, no fragmentation ---
stopifnot(nrow(site_chemical_wide) == n_distinct(filtered$EPA.ID))


# --- 5. Binary presence/absence matrix for clustering (0/1) ---
chem_matrix <- as.matrix(site_chemical_wide[, target_cas])


# --- 6. Yes/No labeled wide table for review/output ---
site_chemical_wide_labels <- site_chemical_wide %>%
  mutate(across(all_of(target_cas), ~ ifelse(. == 1L, "Yes", "No"))) %>%
  rename_with(~ cas_names[.x], all_of(target_cas))

write.csv(site_chemical_wide_labels, file = "site_chemical_profiles_wide.csv", row.names
= FALSE)


# --- 7. Compute Jaccard distance on the binary chemical profile ---
jaccard_dist <- proxy::dist(chem_matrix, method = "Jaccard")


# --- 8. Hierarchical clustering ---
hc <- hclust(jaccard_dist, method = "average")

# inspect the dendrogram to choose k before finalizing
plot(hc, labels = FALSE, main = "Dendrogram of NPL Site Chemical Profiles (Jaccard
distance)")

k <- 10   # number of clusters; adjust after inspecting the dendrogram above
site_chemical_wide$cluster <- cutree(hc, k = k)

# are all the clusters assigned to at least a site?
length(unique(site_chemical_wide$cluster))


# --- 9. Confirm cluster sizes reconcile with the total site count ---
cluster_sizes <- site_chemical_wide %>% count(cluster, name = "n_sites")
```

```r
stopifnot(sum(cluster_sizes$n_sites) == nrow(site_chemical_wide))

write.csv(
  site_chemical_wide %>%
    mutate(across(all_of(target_cas), ~ ifelse(. == 1L, "Yes", "No"))) %>%
    rename_with(~ cas_names[.x], all_of(target_cas)),
  file = "site_chemical_profiles_with_clusters.csv", row.names = FALSE
)
write.csv(cluster_sizes, file = "cluster_size_summary.csv", row.names = FALSE)


# --- 10. Compute per-cluster chemical prevalence (proportion of sites) ---
cluster_prevalence <- site_chemical_wide %>%
  group_by(cluster) %>%
  summarize(across(all_of(target_cas), mean), n_sites = n(), .groups = "drop") %>%
  mutate(cluster_label = paste0("Cluster ", cluster, " (n=", n_sites, ")")) %>%
  select(cluster_label, all_of(target_cas)) %>%
  rename_with(~ cas_names[.x], all_of(target_cas))


# --- 11. Heatmap: cluster (y-axis) x chemical (x-axis), fill = proportion of sites ---
heatmap_data <- cluster_prevalence %>%
  pivot_longer(cols = -cluster_label, names_to = "chemical", values_to = "proportion")

heatmap_plot <- ggplot(heatmap_data, aes(x = chemical, y = cluster_label, fill =
proportion)) +
  geom_tile(color = "white") +
  scale_fill_gradient(low = "white", high = "firebrick", limits = c(0, 1),
                      labels = scales::percent) +
  labs(x = "Nitrogenous organic chemical", y = "Cluster",
       fill = "Proportion\nof sites",
       title = "Nitrogenous Organic Chemical Mixtures by Cluster") +
  theme_minimal() +
  theme(axis.text.x = element_text(angle = 45, hjust = 1))

heatmap_plot

ggsave(filename = "noc_cluster_heatmap.png", plot = heatmap_plot, width = 10, height = 6,
dpi = 300)


# --- 12. Descriptive cluster labels, assigned after reviewing the heatmap above ---
# Named by the dominant/defining chemical(s) characterizing each cluster's
# site profile (Section 2.8); "-only" clusters are sites where that chemical
# is essentially the sole nitrogenous organic COC, "-mix" clusters are sites
# where that chemical co-occurs with others.
cluster_names <- c(
  "1"  = "N-Nitrosodiphenylamine-mix",
  "2"  = "Benzidine-mix",
  "3"  = "Carbazole-mix",
  "4"  = "N-Nitrosodi-n-propylamine-mix",
  "5"  = "NDMA-mix",
  "6"  = "Nitrobenzene-mix",
```

```r
  "7"  = "Aniline-mix",
  "8"  = "Diphenylamine-only",
  "9"  = "4-Nitroaniline-only",
  "10" = "1,2-Diphenylhydrazine-only"
)

cluster_prevalence_labeled <- site_chemical_wide %>%
  group_by(cluster) %>%
  summarize(across(all_of(target_cas), mean), n_sites = n(), .groups = "drop") %>%
  mutate(cluster_label = paste0(cluster_names[as.character(cluster)], " (n=", n_sites,
")")) %>%
  arrange(n_sites) %>%   # ascending, so largest clusters plot at the top of the heatmap
  mutate(cluster_label = factor(cluster_label, levels = cluster_label)) %>%
  select(cluster_label, all_of(target_cas)) %>%
  rename_with(~ cas_names[.x], all_of(target_cas))


# --- 13. Heatmap with descriptive cluster labels on the y-axis ---
heatmap_data_labeled <- cluster_prevalence_labeled %>%
  pivot_longer(cols = -cluster_label, names_to = "chemical", values_to = "proportion")

heatmap_plot_labeled <- ggplot(heatmap_data_labeled, aes(x = chemical, y = cluster_label,
fill = proportion)) +
  geom_tile(color = "white") +
  scale_fill_gradient(low = "white", high = "firebrick", limits = c(0, 1),
                      labels = scales::percent) +
  labs(x = "Nitrogenous organic chemical", y = "Cluster",
       fill = "Proportion\nof sites",
       title = "Nitrogenous Organic Chemical Mixtures by Cluster") +
  theme_minimal() +
  theme(axis.text.x = element_text(angle = 45, hjust = 1))

heatmap_plot_labeled

ggsave(filename = "noc_cluster_heatmap_labeled.png", plot = heatmap_plot_labeled, width =
10, height = 6, dpi = 300)
```